\documentclass[aps,superscriptaddress,showpacs,nofootinbib,twocolumn]{revtex4}

\usepackage{amsmath}
\usepackage{slashed}
\usepackage[dvips]{color}                 
\usepackage{graphicx}
\usepackage{eucal}

\usepackage{mathbbol}

\newcommand{\beq}{\begin{equation}}
\newcommand{\eeq}{\end{equation}}
\newcommand{\bea}{\begin{eqnarray}}
\newcommand{\eea}{\end{eqnarray}}
\newcommand{\nn}{\nonumber}

\usepackage{epsfig}

\begin{document}

\title{Non-perturbative corrections to the quasiparticle velocity in graphene}

\author{Heron Caldas} \email{hcaldas@ufsj.edu.br} \affiliation{Departamento de
  Ci\^{e}ncias Naturais, Universidade Federal de S\~{a}o Jo\~{a}o del Rei,\\
  36301-160, S\~{a}o Jo\~{a}o del Rei, MG, Brazil}

\begin{abstract}
Relativistic fermionic systems have physical quantities calculated by well stablished quantum electrodynamic prescriptions. In the last few years there has been an enormous interest in condensed matter systems in which the fermions exhibit relativistic dispersion, as Dirac fermions in graphene. We employ a non-perturbative method in order to obtain a non-perturbative correction to the quasiparticle velocity in graphene, and compare with the experimental data. We find a better agreement between the quasiparticle velocity corrected with non-perturbative corrections and measurements, when compared with the standard one-loop result. We also investigate the behavior of the beta function of the renormalization group theory, and find that the non-perturbative corrections do not alter the stability of the infrared fixed point found by the standard result.

\end{abstract}

\pacs{11.10.Kk,12.20.-m,68.65.Pq}

\maketitle

\section{Introduction}

In condensed matter (CM) bare electrons are dressed with interactions with the many-body system, forming a ``quasiparticle''~\cite{Mattuck}. A very interesting CM quasiparticle example happens in graphene, a promising material with extraordinary electrical, thermal and optical properties~\cite{intro1,intro2,intro3,intro4,intro5,intro6,intro7}. In graphene charge carriers behave like massless relativistic particles with a conical energy spectrum $E = \hbar  v_F k$, where $\hbar$ is the Planck's constant divided by $2\pi$, $v_F$ is the Fermi velocity, which plays the role of the effective speed of light and $k$ is the wave vector. Electron-electron interactions in graphene lead to a logarithmically divergent correction to the Fermi velocity~\cite{Elias}.

As is well known, the naive perturbative expansion in powers of the coupling constant breaks down in some situations as, for instance, in theories with spontaneously broken symmetries requiring {\it non-perturbative} resummation schemes to obtain reliable results~\cite{T1,T2,T3}. A second case happens in asymptotically free theories where one is not always able to do perturbation theory, since the coupling depends on the energy scale, and at low energies the interaction becomes strong~\cite{MMarino}. Finally, another example which is directly related with this work is the difficulty in conciliate the expansion of quasiparticle velocity $v_F(k)$ in powers of $\alpha$, the effective fine structure constant for graphene (since $\alpha = e^2/\hbar \kappa v_F \simeq 2.2$ i.e., it is greater than unity), with the experimental results~\cite{Sharma}. Here $\kappa$ is the dielectric constant arising from the surrounding medium.  $\alpha = 2.2$ is the maximum allowed value of the bare coupling strength in graphene. This value is found with $\kappa = 1$ which is for graphene in vacuum, and corresponds, therefore, to the most strongly interacting graphene system~\cite{Das}. 

As pointed out in Ref.~\cite{Das2}, experiments detect a quasiparticle velocity enhancement close to the Dirac point, as reported by Elias {\it et al.}~\cite{Elias}, who observed a quasiparticle velocity enhancement on free-standing graphene, while perturbative renormalization group to second order shows that an unphysical fixed point in the Fermi velocity beta function appears at a critical coupling of $\alpha^* = 0.78$~\cite{Das}, indicating a decrease in the quasiparticle (Fermi) velocity $v_F(k)$ at low density for suspended graphene.

Motivated by these examples, in this work we investigate the effects of non-perturbative corrections to the quasiparticle velocity in a two-dimensional Dirac system. Our main goal is to calculate the zero temperature electron self-energy correction to the quasiparticle velocity $v_F$ in graphene in 2D (two space dimensions). We find that the second-order (in the non-perturbative calculations) results agree better with the experimental measurements than the standard (purely one-loop) result.

This paper is organized as follows. In section \ref{section-relat} we calculate the corrections to the Fermi velocity of relativistic-like quasiparticles in 2D. In this section we also study the behavior of the beta function of the renormalization group theory, and find that the non-perturbative corrections do not modify the stability of the infrared fixed point found by the standard (one-loop) result. We conclude in section \ref{conclusions}.

\section{The Non-perturbative Fermion Self-Energy: The MSCR}
\label{section-relat}

As discussed at the Introduction, it is of great importance the development of (alternative) theoretical methodologies, as non-perturbative calculations, that allow the obtention of physical quantities, principally in the situations where perturbation theory ceases to work or are not applicable.

Here we apply the Modified Self-Consistent Resummation (MSCR) procedure in order to implement a resummation of infinity sub-set of diagrams~\cite{Caldas1,Caldas2,Caldas3}. The method consists in calculating the quasiparticle {\it effective} mass through consecutive iterations, assuring the mass renormalization in each step. 

The divergences from the integrals computed in a present step are exactly canceled by the counter-terms that are necessary to be introduced at the previous order of the recalculation (resummation) of the self-energy to carry the renormalization. This renormalization method has been developed, originally in the context of low energy quantum chromodynamics (QCD), in Refs.~\cite{Caldas1,Caldas2,Caldas3} for the resummation method employed here and has been given the name MSCR. For the purposes of this work, we initially concentrate on the full expression for $v_F(k)$ and refer the reader to the above cited works for a detailed treatment of the renormalization procedure.

\subsubsection{Renormalization of the Quasiparticle Velocity in Graphene}

Graphene is a material made of a one-atom-thick sheet of carbon, being a system where the electrons have conical valence and conduction bands, therefore behaving as massless Dirac fermions at low energies~\cite{intro1,intro2,intro3}. Regarding the electron interactions in graphene, the ratio of the strength of the {\it Coulombic} interactions to the analogous quantity in QED is $\alpha_{graphene}/\alpha_{QED} \simeq 300$~\cite{effective}. Because of this graphene can be considered a strong-coupling massless QED in a plane~\cite{intro4}.

To calculate the one-loop renormalization of the Fermi velocity we follow closely Ref.~\cite{Das}, which works with the Euclidean action,

\bea
\label{Son}
S\!\!\!&=&\!\!\!{-}\sum_{a=1}^N\int dt d^2x({\bar\psi}_a\gamma^0\partial_0\psi_a{+}v_F{\bar\psi}_a\gamma^i\partial_i\psi_a{+}A_0{\bar\psi}_a\gamma^0\psi_a)
\nn\\&&\!\!\!+{1\over2g^2}\int dtd^3x(\partial_iA_0)^2.
\eea
The action $S$ above was first used in Ref.~\cite{Son} to describe $2+1$ dimensional four-component massless Dirac fermions interacting through a $3$ dimensional instantaneous Coulomb interaction. In Eq.~(\ref{Son}), $v_F \approx c/300$ is the bare Fermi velocity of the Dirac quasiparticles in graphene. The fields ${\bar\psi}_a$ are four-component fermion fields describing electrons and holes, with $a$ labeling the fermion species. The number of species, $N$, is equal to 2 in real graphene, corresponding to the spin degeneracy. The $\gamma$'s are Dirac matrices satisfying the Euclidean Clifford algebra, $\{\gamma^\mu,\gamma^\nu\}=2\delta^{\mu\nu}$, which can be chosen as

\beq
\gamma^0=\sigma_3\otimes\sigma_3, \qquad \gamma^i=\sigma_i\otimes I,
\eeq
where the $\sigma_i$ are Pauli matrices. 

The coupling $g^2$ is given by
\beq
g^2={2\over1+\epsilon}{e^2\over\epsilon_0}=\frac{4\pi e^2}{\kappa},
\label{coupling-g}
\eeq
where $e$ is the electric charge, $\epsilon_0$ is the vacuum permeability, and $\epsilon$ and $\kappa$ are two different definitions of the dielectric constant of the substrate (SI and cgs units, respectively). Associating these quantities with the Fermi velocity, one can define,

\beq
\alpha\equiv\frac{g^2}{4\pi v_F}=\frac{e^2}{2\pi(1+\epsilon)\epsilon_0 v_F}=\frac{e^2}{\kappa v_F}.\label{defofalpha}
\eeq
It is appropriate to make use of the quasirelativistic notation,
\beq
\slashed p=\gamma^0p_0+v_F\vec{\gamma}\cdot\vec{p},\qquad p^2=p_0^2+v_F^2|\vec{p}|^2.
\eeq
The propagator of a massless free fermion is given by

\beq
G_0(p)=\frac{i}{\slashed p}=\frac{i\slashed p}{p^2},
\eeq
while the effective propagator for the Coulomb interaction reads
\beq
D_0(p)=g^2\int\frac{dp_z}{2\pi}\frac{1}{p_z^2+|\vec{p}|^2}=\frac{g^2}{2|\vec{p}|},
\eeq
and the interaction vertex is $i\gamma^0$. Every closed fermion loop contributes an overall minus sign to the value of the diagram.

The one-loop diagram has been evaluated several times in the literature, and following Ref.~\cite{Das} in Appendix~\ref{ApA}, we find

\beq
\Sigma_1(q)=\frac{i g^2}{16\pi}\vec{q}\cdot\vec\gamma\log(\Lambda/|\vec{q}|),
\label{selfenergyGrapheno}
\eeq
where $\Lambda$ is an ultraviolet cut-off. For the case of graphene, the physical cut-off is $\Lambda \approx 1/a$, with $a$ being the graphene lattice constant, and represents the momentum scale up to which the spectrum is Dirac-like~\cite{Kotov}.

Since the full fermion propagator is given by
\beq
G(p)=\frac{i}{\slashed p-i\Sigma(p)},
\label{full}
\eeq
the one-loop self-energy leads to a renormalization of the quasiparticle velocity,
\bea
v_q&\equiv&v_F^*(q)\nn\\
&=&v_F+\frac{g^2}{16\pi}\log(\Lambda/|\vec{q}|)\nn\\
&=&v_F\left[1+\frac{\alpha}{4}\log(\Lambda/|\vec{q}|)\right],
\label{vel1}
\eea
meaning that the quasiparticle velocity increases, since $\alpha>0$ and $\Lambda/|\vec{q}| \gg 1$.

With the aid of Eq.~(\ref{coupling-g}), and with $v_q$ in Eq.~(\ref{vel1}) an equation for the coupling as a function of the momentum and the bare coupling can be obtained

\bea
\frac{\alpha_q}{\alpha}=\frac{1}{1+\frac{\alpha}{4}\log(\Lambda/|\vec{q}|)} = \frac{v_F}{v_q}.
\label{alphaq}
\eea
in order to eliminate the arbitrary ultraviolet momentum cutoff $\Lambda$ can find~\cite{Das2} expressions akin to those above in another momentum, say $k$ 

\bea
\frac{\alpha_k}{\alpha}=\frac{1}{1+\frac{\alpha}{4}\log(\Lambda/|\vec{k}|)} = \frac{v_F}{v_k}.
\label{alphak}
\eea
Instead of defining a velocity ratio as in Ref.~\cite{Das}, which still depends on $\Lambda$ through $\alpha_k$ ($\frac{v_q}{v_k}=1+\frac{\alpha_k}{4}\log(|\vec{k}|/|\vec{q}|)$, with $\alpha_k$ given by Eq.~(\ref{alphak})), we define a velocity difference, which is independent of $\Lambda$,

\bea
\frac{\alpha}{\alpha_q} - \frac{\alpha}{\alpha_k} =\frac{v_q-v_k}{v_F}=\frac{\alpha}{4}\log(|\vec{k}|/|\vec{q}|),
\label{vqk}
\eea
which relates the difference between the physical velocities at momentum $|\vec{q}|$ and $|\vec{k}|$, respectively, with no cutoff $\Lambda$ dependence.

\subsubsection{One-loop Renormalization Group Analysis}

Let us now apply renormalization group (RG) theory by considering the effect of changing the momentum scale $q=|\vec{q}|$ of $v_q$ in Eq.~(\ref{vel1}). From Eq.~(\ref{vqk}) we can obtain the beta function for the effective coupling, by means of the renormalization group equation

\bea
\beta(\bar \alpha)=k \frac{\partial \bar \alpha}{\partial k}=-\frac{1}{4}\frac{1}{\left(\frac{1}{4}\log(k/q) \right)^2},
\label{beta1}
\eea
where $k=|\vec{k}|$, and from Eq.~(\ref{vqk}) we have defined $\frac{1}{\bar \alpha} \equiv \frac{1}{ \alpha_q} - \frac{1}{\alpha_k}$. Integration of the above $\beta(\bar\alpha)$ equation gives

\bea
\frac{1}{\bar \alpha(k)} - \frac{1}{\bar \alpha(q)}= \frac{1}{4} \ln \left( \frac{k}{q} \right),
\label{beta2}
\eea
which shows that the coupling strength $\bar \alpha(k)$ decreases with increasing $k$.

Notice that from Eq.~(\ref{beta1}) we have

\bea
\beta(\bar \alpha)=-\frac{1}{4} \bar \alpha^2.
\label{beta3}
\eea
This result tells us that a (trivial) fixed point (FP) occurs at $\bar \alpha = \bar\alpha_c =0$. This FP is stable, since

\bea
\frac{d \beta(\bar \alpha)}{d \bar\alpha}|_{\bar\alpha = \bar\alpha_c}=0.
\label{beta4}
\eea

This stable FP corresponds to an infrared (IR) point, since it is reached through the limit $q \to 0$~\cite{Kotov}.

\begin{figure}[htb]
  \vspace{0.1cm}
  \epsfig{figure=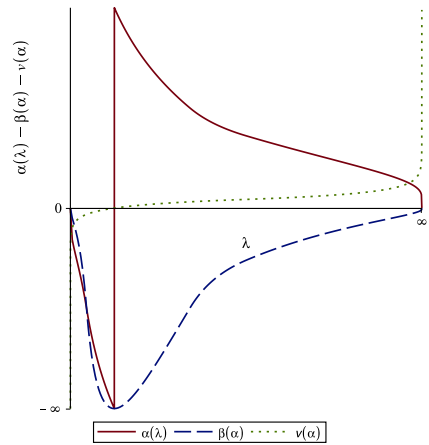,angle=0,width=6cm}
\caption[]{\label{Beta-Function} Behavior of the one-loop coupling strength $\bar\alpha$, the $\beta(\bar\alpha)$ function and the renormalized Fermi velocity $v \propto 1/\bar\alpha$, as a function of $\lambda= k/q$.}
\end{figure}

In Fig.~\ref{Beta-Function} we show the behavior of the coupling strength $\bar\alpha$, and the $\beta(\bar\alpha)$ function from Eqs.~(\ref{beta2}) and~(\ref{beta3}), respectively, as a function of $\lambda\equiv k/q$. Both functions vanish at the limit $\lambda \to \infty$. On the other hand, this also means that the renormalized quasiparticle velocity $v \equiv (v_{q}-v_{k})/v_F \propto \bar\alpha^{-1}$ diverges in this limit. The renormalized quasiparticle velocity $v(\bar\alpha)$ is also shown in Fig.~\ref{Beta-Function}. The functions $\alpha(\lambda_{UV})$ and $\beta(\bar\alpha(\lambda_{UV}))$ diverge at the ultraviolet point $\lambda=\lambda_{UV}=1$. At this UV point $v$ is zero. Lastly, at the (unattainable) point $\lambda=0$, $\bar\alpha(\lambda=0)=\beta(\bar\alpha(\lambda=0))=0$, and $v$ diverges to $-\infty$.

\subsubsection{The Non-perturbative Electron Quasiparticle Velocity in Graphene}

We now obtain the non-perturbative corrections to the renormalization of the quasiparticle velocity we have just obtained up to second-order in the iterations. 

As is well known, owing to the massless character of electrons in the model described by Eq.~(\ref{Son}), it is sensitive to retardation effects of the electromagnetic propagation~\cite{GonzalezNPB94}. In static models like this, with an instantaneous photon propagator i.e., corresponding to $c=\infty$, the Fermi velocity grows without bound (in opposition to what happens in the retarded - corresponding to $c$ finite - model~\cite{GonzalezNPB94}) in the infrared~\cite{V-Guinea}. Thus, the use of the non-retarded action $S$ (with such an instantaneous long range Coulomb interaction) in Eq.~(\ref{Son}) deserves some comments on it. As pointed out in Ref.~\cite{GonzalezNPB94}, one should not expect any discrepancy in the computation of local quantities, i.e., cutoff dependent ones, in a non-retarded and in a retarded model in the limit $v_F/c \to 0$. However, different results should be noticeable in the computation of nonlocal quantities, due to the dissimilar forms of the interactions in the infrared regime~\cite{GonzalezNPB94}. In addition, because of the mentioned infrared instabilities, perturbation theory looses its predictive power and other methods (such as the non-perturbative MSCR) may be employed. There are some particular cases in which retardation effects are not important as, for instance, that of Van der Waals interactions. In fact, it was demonstrated in Ref.~\cite{Gomes} that retardation effects has little significance in undoped graphene, for both zero and finite temperature. It is also worth to mention that in the investigation of finite-temperature Casimir effects for graphene, the transition from the retarded regime to a non-retarded one, happens at separations characterized by a material wavelength~\cite{Ignat}.

Proceeding as before, we define $v_F^*(q)$ in Eq.~(\ref{vel1}) as $v_{F1}(q)$, and obtain $\Sigma_2 = \Sigma_2(v_{F1}(q))$,

\beq
\Sigma_2(v_{F1}(q))=\frac{i \alpha}{4} v_{F1}(q) \vec{q}\cdot\vec\gamma\log(\Lambda/|\vec{q}|).
\label{selfGrapheno2}
\eeq
Inserting the above equation into Eq.~(\ref{full}), we obtain

\bea
v_{F2}(q) &=&v_F+v_{F1}(q)\frac{\alpha}{4}\log(\Lambda/|\vec{q}|) \to \nn\\
\frac{v_{F2}(q)}{v_{F}}&=&1+  \frac{\alpha}{4}\log(\Lambda/|\vec{q}|) + \left( \frac{\alpha}{4} \right)^2\left[  \log(\Lambda/|\vec{q}|)\right]^2,
\label{vel2}
\eea
where we have made use of Eq.~(\ref{vel1}). The MSCR is somewhat very close to the DuBois's approach, i.e., an {\it iteration-perturbation method}, which generates a series of iterations~\cite{Dubois}.

\begin{figure}[htb]
  \vspace{0.1cm}
  \epsfig{figure=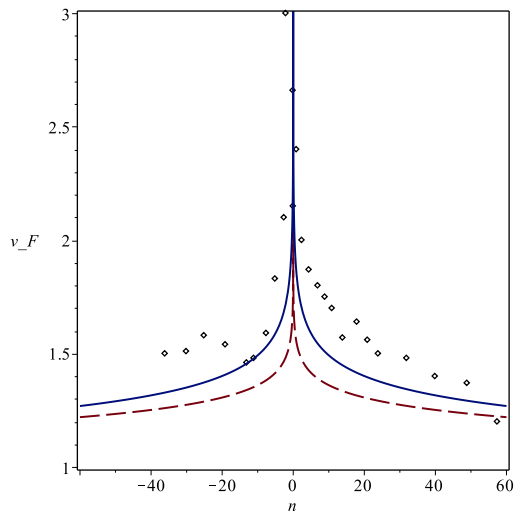,angle=0,width=7cm}
\caption[]{\label{Fermi-Velocity} Behavior of the Fermi velocity $v_{F1}(n)$, dashed curve, and $v_{F2}(n)$, solid curve, (plotted in units of $10^{6}ms^{-1}$), as a function of carrier density $n$ (in units of $10^{12}cm^{-2}$). The experimental data points are taken from Ref.~\cite{Elias}.}
\end{figure}

The behavior of $v_{F2}$ as a function of the density $n$ can be investigated by noticing that the Fermi momentum is proportional to the square root of the carrier density, $q_F=(\pi n)^{1/2}$, while the momentum cutoff $\Lambda=(\pi n_c)^{1/2}$, with a fixed reference density $n_c=10^{15}cm^{-2}$~\cite{Das}. Then we find

\bea
\frac{v_{F2}(n)}{v_{F}}&=&1+  \frac{\alpha}{8}\log \left(\frac{n_c}{n} \right) + \frac{1}{4}\left( \frac{\alpha}{4} \right)^2\left[\log \left(\frac{n_c}{n} \right) \right]^2.
\label{vel-n}
\eea
In order to compare $v_{F2}(n)$ obtained above with the experimental measurements of the renormalized Fermi velocity for suspended graphene, we employ the same approximation used in Ref.~\cite{Elias} to take into account the screening by graphene's charge carriers. This approach is implemented by defining an effective screening constant $\epsilon_G$ such that $\alpha \to \alpha/\epsilon_G$ in Eq.~(\ref{vel-n}). The whole range of $n$ in Ref.~\cite{Elias} had a best fit for $\epsilon_G \sim 3.5$. In Fig.~\ref{Fermi-Velocity} we show $v_{F1}(n)$ and $v_{F2}(n)$ for $\alpha=2.2$ and $\epsilon_G \sim 3.5$, where $v_{F1}(n)$ (dashed curve) is the standard one-loop result, while $v_{F2}(n)$, calculated with the second-order of the iteration within MSCR, is the solid curve. $v_{F2}(n)$ shows a much better agreement (as compared to $v_{F1}(n)$) with the experimental data points, obtained in Ref.~\cite{Elias}.

Notice that the renormalized Fermi velocity for graphene deposited on a substrate shows the same (qualitative) behavior as for suspende graphene. See, for instance, Ref.~\cite{Yu}, where, for graphene on a substrate, $\alpha < 2.2$, which corresponds to a dielectric constant $\kappa >1$~\cite{Das}.

As pointed out in Ref.~\cite{Sharma}, it is worth mentioning that the renormalization of the Fermi velocity has been computed to second-order of perturbation theory (in powers of the dimensionless interaction constant) and conflicting results have been found as, for example, in Ref.~\cite{Mish}, it was found

\bea
v_{F}(q) = v_{F}\left\{1+ \left[ \frac{g_0}{4} - g_0^2 \left( \frac{5}{6} - \ln2 \right) \right] \log(\Lambda/|\vec{q}|)\right\},
\label{vel3}
\eea
where $g_0=e^2/\kappa \nu_0$, where $\kappa$ is the dielectric constant of a substrate and $\nu_0$ is the bare electron velocity, then $g_0=\alpha$, and in Ref.~\cite{Das}

\bea
v_{F}(q) = v_{F}\left\{1+ \left[ \frac{\alpha}{4} + \alpha^2 \left( \frac{3 \log2 - 4}{6} \right) \right] \log(\Lambda/|\vec{q}|)\right\}.
\label{vel4}
\eea
Actually, the above result is formed by the first term of a general expression found in Ref.~\cite{Das}

\bea
\frac{v_{F}(q)}{v_{F}}=1+ \sum_{n=1}^{\infty} F_n(\alpha)  \log^n(\Lambda/|\vec{q}|),
\label{vel5}
\eea
where $F_1(\alpha)= f_{1}\alpha+f_{2}\alpha^2$, with $f_1=\frac{1}{4}$, and $f_2= \frac{3 \log2 - 4}{6}$.  Eq.~(\ref{vel5}) is an indicative of breakdown of perturbation theory.

\subsubsection{One-loop plus Non-perturbative Corrections Renormalization Group Analysis}

From Eq.~(\ref{vel2}) we can obtain the momentum-dependent coupling constant 

\bea
\frac{\alpha_{q2}}{\alpha} = \frac{1}{1+  \frac{\alpha}{4}\log(\Lambda/q) + \left( \frac{\alpha}{4} \right)^2\left[  \log(\Lambda/q)\right]^2}.
\label{alpha2-1}
\eea
The difference of the inverse of the coupling at two different momenta, gives

\bea
&&\frac{\alpha}{\alpha_{2q}} -\frac{\alpha}{\alpha_{2k}} = \frac{v_{F2}(q)-v_{F2}(k)}{v_{F}}=\\
\nonumber
&&   \frac{\alpha}{4}\log(k/q) + \left( \frac{\alpha}{4} \right)^2\left[  (\log(\Lambda/q))^2 - (\log(\Lambda/k))^2\right].
\label{alpha2-2}
\eea
Defining $\frac{1}{\bar \alpha_{2}} \equiv \frac{1}{\alpha_{2q}} -\frac{1}{\alpha_{2k}}$ as we did at one-loop, we find a beta function for the effective coupling

\bea
&&\beta(\bar \alpha_{2})=k \frac{\partial \bar \alpha_{2}}{\partial k}\\
\nonumber
&=&-\frac{1}{4} \frac{ \left[1+\frac{\alpha}{2}\log(\Lambda/k)  \right]}{\left[\frac{1}{4}\log(k/q) + \frac{\alpha}{16} \left(  (\log(\Lambda/q))^2 - (\log(\Lambda/k))^2\right) \right]^2} \\
\nonumber
&=&-\frac{1}{4} \left[1+\frac{\alpha}{2}\log(\Lambda/k)  \right] \bar\alpha_{2}^2.
\label{beta2-1}
\eea
As happens at one-loop, a fixed point occurs at $\bar\alpha_{2}=\bar\alpha_{2c}=0$. Thus, up to second-order of the non-perturbative iteration we do not find any fixed point $\bar\alpha_{2} \neq 0$. Again, this is a  stable IR-FP, since

\bea
\frac{d \beta(\bar \alpha_2)}{d \bar\alpha_2}|_{\bar\alpha_2 = \bar\alpha_{2c}}=0.
\label{beta4}
\eea

\section{Summary}
\label{conclusions}

To summarize, we have investigated the effect of non-perturbative corrections to the Fermi velocity in graphene. Interestingly, we find that although the electrons near the Dirac points (i.e., the contact points between the valence and conduction bands) in graphene have the well known {\it Dirac dispersion}, the evolution of their coupling with momentum (energy) scale reveals a behavior of many non-abelian gauge theories, as QCD, possessing a negative $\beta$ function. We find that both the first and second-order results of the non-perturbative calculation shown in Fig.~\ref{Fermi-Velocity} are in qualitative agreement with the experimental data obtained in Ref.~\cite{Elias}. However, the second-order result of the non-perturbative calculation is in quantitative agreement with the measurements of the Fermi velocity renormalization in suspended graphene, taken from Ref.~\cite{Elias}.

We also investigated the behavior of the $\beta$ function of the renormalization group theory, and find that the non-perturbative corrections do not modify the stability of the infrared fixed point found by the standard (one-loop) result.

Our approach might also be useful in the investigation of various physical quantities, in graphene and other similar fermion systems. The study of the non-perturbative corrections to the effective mass and consequently to the pairing gap of a 1d fermion system is in progress and will be presented elsewhere.
\vspace{0.5cm}

\section{Acknowledgments}

We would like to thank E. Barnes and J. Hofmann for helpful discussions. The author acknowledges partial support by CNPq-Brazil and FAPEMIG.
\vspace{0.1cm}

\appendix

\section{Explicit calculation of the one-loop electron self-energy in graphene}
\label{ApA}

The one-loop electron self-energy diagram was originally evaluated in Ref.~\cite{GonzalezNPB94}, however, we follow the steps from Ref.~\cite{Das}. The diagram is written as

\bea
\Sigma_1(q)&=&-\int\frac{d^3k}{(2\pi)^3}\gamma^0 G_0(k+q)\gamma^0 D_0(k)
\nn\\&=&-\frac{ig^2}{2}\int\frac{d^3k}{(2\pi)^3}\gamma^0\frac{\slashed k+\slashed q}{(k+q)^2}\gamma^0\frac{1}{|\vec{k}|}
\nn\\&=&-\frac{ig^2}{2}\int \frac{d^2k}{(2\pi)^2}\frac{1}{|\vec{k}|}\gamma^0 K(q,\vec{k})\gamma^0,
\eea
\bea
K(q,\vec{k})&=&\int\frac{dk_0}{2\pi}\frac{\slashed k+\slashed q}{(k+q)^2}\nn\\&=&\int\frac{dk_0}{2\pi}\left[\frac{v_F(\vec{k}+\vec{q})\cdot\vec{\gamma}}{k_0^2+v_F^2|\vec{k}+\vec{q}|^2}+\frac{k_0\gamma^0}{k_0^2+v_F^2|\vec{k}+\vec{q}|^2}\right]
\nn\\&=&\frac{(\vec{k}+\vec{q})\cdot\vec{\gamma}}{2|\vec{k}+\vec{q}|},\label{expressionforY}
\eea
\beq
\Sigma_1(q)=\frac{ig^2}{4}\int\frac{d^2k}{(2\pi)^2}\frac{1}{|\vec{k}|}\frac{(\vec{k}+\vec{q})\cdot\vec{\gamma}}{|\vec{k}+\vec{q}|}.
\eeq
In order to perform the remaining integrals, we choose the coordinate system such that $\vec{q}=(|\vec{q}|,0)$. As in Ref.~\cite{Das}, we adopt the transformation to elliptical coordinates
\bea
&&k_x=\frac{|\vec{q}|}{2}(\cosh\mu\cos\nu-1),\qquad k_y=\frac{|\vec{q}|}{2}\sinh\mu\sin\nu,\nn\\&& d^2k=\frac{|\vec{q}|^2}{4}(\cosh^2\mu-\cos^2\nu)d\mu d\nu,
\eea
yielding
\bea
&&\!\!\!\!\!\!\Sigma_1(q)=
\nn\\&&\!\!\!\!\!\!\frac{i g^2|\vec{q}|}{32\pi^2}\gamma\cdot\int_0^{2\pi}d\nu\int_0^{\mu_{max}}d\mu(1+\cos\nu\cosh\mu,\sin\nu\sinh\mu).\nn\\&&\label{eqn12}
\eea
The integral has been regulated by including the cutoff, $\mu_{max}$. We would like to relate this cutoff to a more physical cutoff on the momentum: $\Lambda\ge|\vec{k}|$. The mapping to elliptical coordinates given above implies
\beq
|\vec{k}|=\frac{|\vec{q}|}{2}(\cosh\mu-\cos\nu),
\eeq
so that
\bea
&&\Lambda=\frac{|\vec{q}|}{2}(\cosh\mu_{max}-\cos\nu)\nn\\&&\Rightarrow\mu_{max}=\cosh^{-1}\left(\frac{2\Lambda}{|\vec{q}|}+\cos\nu\right).
\eea
The three integrals in (\ref{eqn12}) then evaluate to
\bea
&&\!\!\!\!\!\!{\int_0^{2\pi}}d\nu{\int_0^{\mu_{max}}}d\mu={\int_0^{2\pi}}d\nu \cosh^{{-}1}\left(\frac{2\Lambda}{|\vec{q}|}{+}\cos\nu\right)
\nn\\&&=2\pi\log(4\Lambda/|\vec{q}|){+}O(|\vec{q}|^2/\Lambda^2),\nn\\
&&\!\!\!\!\!\!{\int_0^{2\pi}}d\nu\cos\nu{\int_0^{\mu_{max}}}d\mu \cosh\mu
\nn\\&&={\int_0^{2\pi}}d\nu\cos\nu\sqrt{\left(\frac{2\Lambda}{|\vec{q}|}{+}\cos\nu\right)^2{-}1}=\pi{+}O(|\vec{q}|^2/\Lambda^2),\nn\\
&&\!\!\!\!\!\!{\int_0^{2\pi}}d\nu\sin\nu{\int_0^{\mu_{max}}}d\mu \sinh\mu
\nn\\&&={\int_0^{2\pi}}d\nu\sin\nu\left(\frac{2\Lambda}{|\vec{q}|}{+}\cos\nu\right)=0.
\eea
Plugging these results into (\ref{eqn12}), and reverting to a general coordinate system, i.e., $\vec{q}=(q_x,q_y)$, one finally obtains
\bea
\Sigma_1(q)\!\!\!&{=}&\!\!\!\frac{i g^2|\vec{q}|}{32\pi^2}\gamma^1\left[2\pi\log(\Lambda/|\vec{q}|){+}4\pi\log2{+}\pi\right]{+}O\left(\frac{|\vec{q}|^2}{\Lambda^2}\right)
\nn\\\!\!\!&\rightarrow&\!\!\!\frac{i g^2}{16\pi}\vec{q}\cdot\vec\gamma\left[\log(\Lambda/|\vec{q}|){+}2\log2{+}1/2\right]{+}O\left(\frac{|\vec{q}|^2}{\Lambda^2}\right).\nn\\&&
\eea
Then we have,

\beq
\Sigma_1(q)=\frac{i g^2}{16\pi}\vec{q}\cdot\vec\gamma\log(\Lambda/|\vec{q}|),
\label{selfenergyGrap}
\eeq
which is the self-energy in Eq.~(\ref{selfenergyGrapheno}). In the above equation the finite part ($2\log2 + 1/2$) has been absorbed into a redefinition of the ultraviolet momentum cutoff $\Lambda$.

\end{document}